\documentclass[%
 reprint,
longbibliography,
 amsmath,amssymb,
 aps,
]{revtex4-1}

\usepackage{graphicx}
\usepackage{dcolumn}
\usepackage{bm}
\usepackage{subfigure}
\usepackage{caption}
\usepackage[justification=justified,singlelinecheck=false]{caption}
\usepackage{color}

\preprint{APS/123-QED}
\begin{document}
\title{Frustration-induced supersolid phases of extended Bose-Hubbard model in the hard-core limit}

\author{Wei-Lin Tu$^{1}$, Huan-Kuang Wu$^{2}$, and Takafumi Suzuki$^{3}$}

\affiliation{
$^{1}$\textit{Institute for Solid State Physics, University of Tokyo, Kashiwa, Chiba 277-8581, Japan}\\ 
$^{2}$\textit{Department of Physics, Condensed Matter Theory Center and Joint Quantum Institute, University of Maryland, College Park, MD 20742, USA}\\
$^{3}$\textit{Graduate School of Engineering, University of Hyogo, Hyogo, Himeji 670-2280, Japan}
}

\date{\today}

\begin{abstract}
We investigate exotic supersolid phases in the extended Bose-Hubbard model with infinite projected entangled-pair state, numerical exact diagonalization, and mean-field theory. We demonstrate that many different supersolid phases can be generated by changing signs of hopping terms, and the interactions along with the frustration of hopping terms are important to stabilize those supersolid states. We argue the effect of frustration introduced by the competition of hopping terms in the supersolid phases from the mean-field point of view. This helps to give a clearer picture of the background mechanism for underlying superfluid/supersolid states to be formed. With this knowledge, we predict and realize the $d$-wave superfluid, which shares the same pairing symmetry with high-$T_c$ materials, and its extended phases. We believe that our results contribute to preliminary understanding for desired target phases in the real-world experimental systems.
\end{abstract}

\pacs{Valid PACS appear here}
\maketitle

\section{\label{sec:level1}Introduction}

Supersolid (SS) phase, after its first proposition by Penrose and Onsager \cite{Penrose}, has attracted much attention from both experimental and theoretical aspects \cite{Boninsegni}. The nature of the SS state is characterized by the coexistence of crystal order and superfluidity, namely the coexistence of diagonal and off-diagonal long-range order. The SS phase is formed by adding dopants at a commensurate filling, where perfect crystal exists. These dopants then condensate and contribute to the superflow \cite{Andreev, Chester, Leggett} while keeping the crystal order. This scenario is called the ``defect-condensation" \cite{ChenSS}. 

Although great efforts have been paid searching for the SS phase experimentally, the observation of its exotic property has been never reported yet in natural materials. 
In 2004, Kim and Chan have reported the coexistence of solidity and superfluidity in helium-4 \cite{Kim}. 
They have observed a sudden drop of the non-classical rotational moment of inertia in the torsional oscillator experiment. 
After their work, many experiments have been performed in order to argue more of this fascinating phenomenon \cite{Kim2, Rittner, Day}.
However, it has later been demonstrated that the supersolidity is lost when the effect of elastic properties of helium is removed by careful experiments \cite{Day2, Kim3}.

With the observation of supersolidity from helium seeming to be unlikely, our attention turns to another platform, the ultracold atomic gases \cite{Bloch, Bloch2, Windpassinger, Tomza}.
Many possible scenarios to realize the SS state have been proposed for the ultracold atomic gases by adding additional interactions. 
Among all, the dipolar gases \cite{Lahaye, Dutta} seem to become a more reliable platform for supersolidity. 
In quite recent experiments, strong evidences of supersolidity have been found in dipolar atoms made of erbium and dysprosium \cite{Tanzi, Bottcher, Chomaz, Natale, Tanzi2, GuoS}.

With the advanced techniques introduced by ultracold experiments, theorists have also been urged to investigate this issue more deeply. 
One of the most commonly used Hamiltonians for tackling this issue is the extended Bose-Hubbard (EBH) model, which can be experimentally realized by the dipolar gases \cite{Lahaye, Dutta, Baier}. 
In earlier works, it has been revealed that by the next-nearest-neighbor interaction, the SS phase can be formed and appears via a second-order phase transition from the superfluid (SF) phase on a square lattice \cite{Batrouni, Hebert}.
When the further long-range interaction, namely the dipole-dipole interaction in the cold atom experiment, is included into the EBH model, the SS phase still survives \cite{Ohgoe}. 
These results have highlighted the importance of a longer-range interaction for the stabilization of the SS state.

In the effect of the long-range interaction, the frustration between interactions, which seems to be essential to realize the quantum spin liquid in many modern condensed matter systems\cite{Savary, ZhouS}, can be crucial in forming the SS phase. 
A direct mapping from hard-core bosons to S=1/2 spins enables us to discuss ordering phases from one side complementary to the other side \cite{ChanS}. 
On the other hand, the abundant underlying physics of fermionic Hubbard models concerning high-$T_c$ superconductivity \cite{Anderson, Corboz, Tu, Zheng, TuS} also implies that there could be some interesting features for the bosonic side.

Early efforts upon the EBH model for searching the SS state were made by mainly the quantum Monte Carlo (QMC) method and contributed to fruitful results \cite{ChenSS, Ng, Dang,Suzuki}. 
In these studies, the systems with the frustrated hopping interactions were ignored owing to the negative sign problem in QMC calculations. 
However, recent works concerning the same model have shown that with the frustration induced by a negative next-nearest-neighbor hopping $t'$, a peculiar SS phase, named after half supersolid (HSS), can be generated \cite{Dong, ChenS}. 
To further investigate the details of HSS, one might need other numerical approaches.

In this paper, our central method for solving the EBH model in the hard-core limit is through one of the tensor network (TN) ansatz, known as infinite projected entangled-pair state (iPEPS) \cite{Jordan}. 
For this TN ansatz in a square lattice, quantum state on each lattice site is represented by a rank-5 tensor with one physical index of dimension $d$ and four auxiliary indices of dimension $D$ in a two-dimensional plane. Because of the hard-core limit, each site can have two different states and therefore $d=2$, while $D$ can be chosen freely but it helps enhance the accuracy with larger numbers. To attain the thermodynamic limit, here we use the corner-transfer-matrix algorithm \cite{Nishino, Orus, Corboz}. After the iterative convergence, we obtain a series of environment tensors surrounding the unit cell, which we choose to be equal to the size of $2 \times 2$. For the corner and edge tensors, composing the environment tensors, their bond dimension, $\chi$, is chosen to be large enough ($\chi(D) > D^2$) for minimizing the error caused by the usage of finite $\chi$.

To obtain a proper ground state of the target Hamiltonian in iPEPS, we can achieve it by projecting the initial states into the imaginary-time evolution or through the variational update \cite{Corboz2, LiaoS}. 
In the following calculation, we apply the imaginary-time evolution in which the environment tensors are not included while all tensors of the unit cell being updated.
Thus, this imaginary-time evolution method is also called the simple update \cite{JiangS}. 
Compared with another more accurate but computationally expensive update algorithm, namely the full-update method \cite{Jordan, Phien}, 
the simple-update method would lose its accuracy, once the system of interest is highly correlated. 
Nevertheless, the simple-update method with small $D$ has provided the accurate phase diagrams with a dimer model \cite{Wessel}, which is in fact more entangled than our model owing to the longer-range interdimer coupling term.
Therefore, we apply simple-update iPEPS as one of our numerical methodologies.
Besides iPEPS, it has been known that the mean-field theory works well for the EBH model \cite{ChenS}.
Thus, we employ the mean-field theory proposed by Matsubara and Matsuda \cite{Matsubara} to compare with the results by iPEPS. 
In addition to the above two methods, we execute exact diagonalization (ED) numerically.
The combination of ED and the finite-size scaling also provides further evidences of the phases we have found.

This paper is organized in the following way. 
In Section II, we will present our results. We firstly define the order parameters for distinguishing different phases and apply the iPEPS and mean-field approaches for constructing phase diagrams of given hopping and interacting amplitudes in Section II A and B. 
We then re-examine the existence of these phases with ED in Section II C. 
In Section II D, we analyze the causes of found phases in Section II B through a mean-field interpretation and construct general rules in pursuit of desired SF/SS states. 
We follow these rules and predict a $d$-wave SF along with its related phases. 
Our conclusion is carried in Section III and the justification of finite-$D$ iPEPS is included in the Appendix.

\section{\label{sec:level1}Results}

\subsection{\label{sec:level2}Hamiltonian and order parameters}

We study the hard-core EBH model on a square lattice. 
The Hamiltonian is written by
\begin{equation}
\begin{aligned}
H=&-t\sum_{\langle i,j \rangle}(b^\dagger_{i}b_{j}+H.C.)-t'\sum_{\langle \langle i,j \rangle \rangle}(b^\dagger_{i}b_{j}+H.C.)\\
&+V_1\sum_{\langle i,j\rangle}n_i n_j+V_2\sum_{\langle \langle i,j \rangle \rangle}n_i n_j -\mu \sum_i n_i ,
\end{aligned}
\label{Hamiltonian}
\end{equation}
where $b^\dagger_{i}$ ($b_{i}$) stands for the creation (annihilation) operator of the hard-core boson and $n_i=b^\dagger_{i}b_{i}$.
$\langle i,j \rangle$ and $\langle \langle i,j \rangle \rangle$ denote the summation for the nearest-neighbor (nn) and next-nearest-neighbor (nnn) pairs, respectively. 
$V_1$ and $V_2$ represent the inter-site interactions and are generally taken to be positive (repulsive). 
Since we consider the hard-core limit, there can be only two possible states, the empty and occupied states, for each lattice site. 

In recent two papers \cite{Dong, ChenS}, the authors have studied the EBH model and found HSS with negative $t'$, which causes the frustration of Hamiltonian. 
This motivates us to wonder if there are other exotic SS or even SF phases once our model is frustrated in a different way, and if we are able to understand/predict the existence of distinct SF/SS phases. 
Therefore, in this work, we strive to provide a larger scope for underlying phases carried by the EBH model (1) in the hard-core limit. 
To properly distinguish phases, we define our order parameters in this subsection. 

\begin{figure}[t]
\centering
\includegraphics[width=0.48 \textwidth]{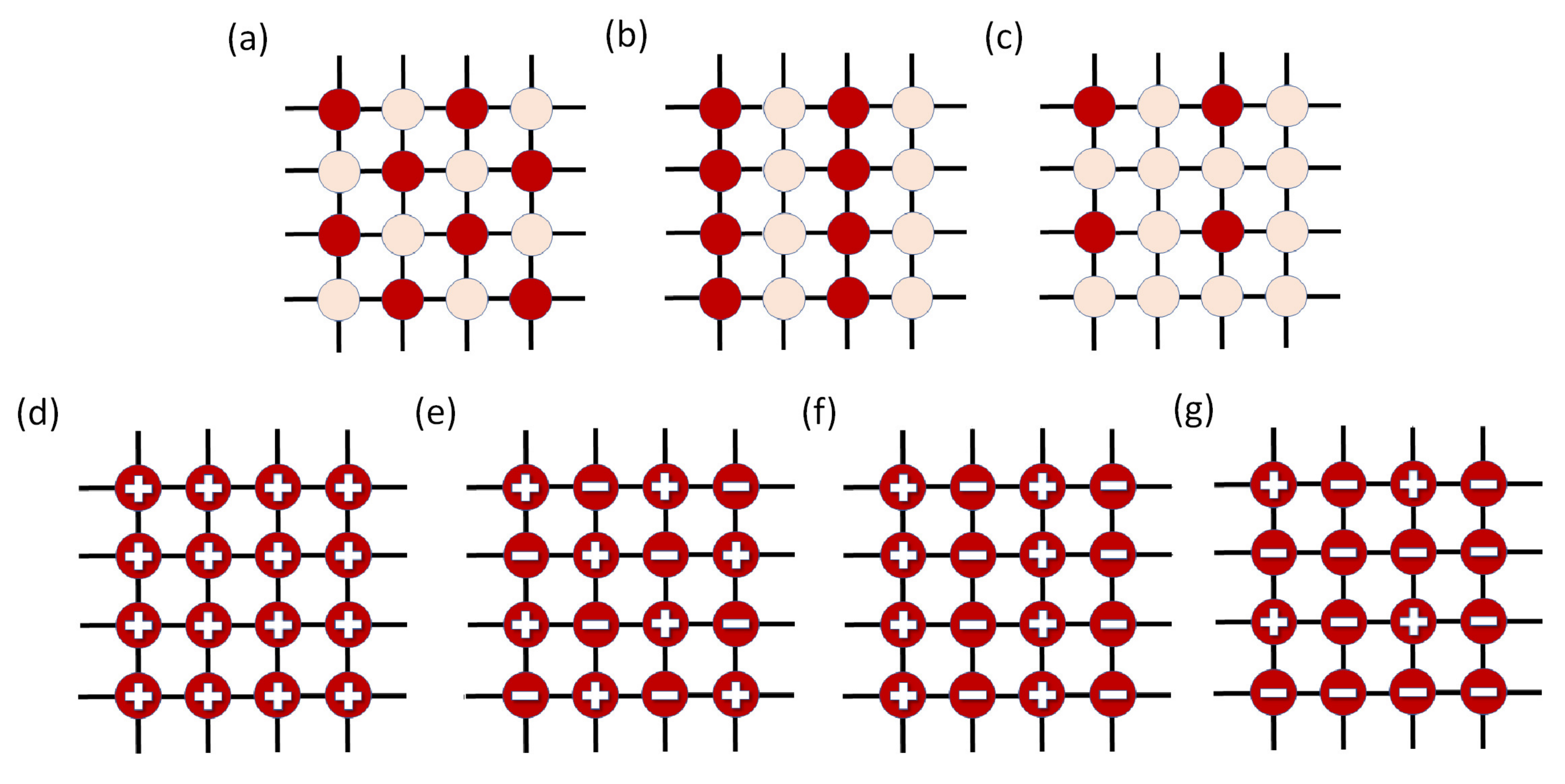}
\caption{\raggedright Configurations for different orders: (a)-(c) are for real-space particle density modulation. They correspond to (a) checkerboard, (b) stripe, and (c) quarter-filled modulations. Sites with darker color are the occupied (largely filled) sites and the others denote empty (lightly filled) sites. (d)-(g) stand for possible patterns of arg[$\langle b_{i} \rangle$] in real space, and we call them as homogeneous, checkerboard, collinear, and star sign patterns, respectively. Notice that each pattern is invariant under a global $Z_2$ transformation.}
\label{Fig.1}
\end{figure} 

\begin{figure*}[t]
\centering
\includegraphics[width=1.0 \textwidth]{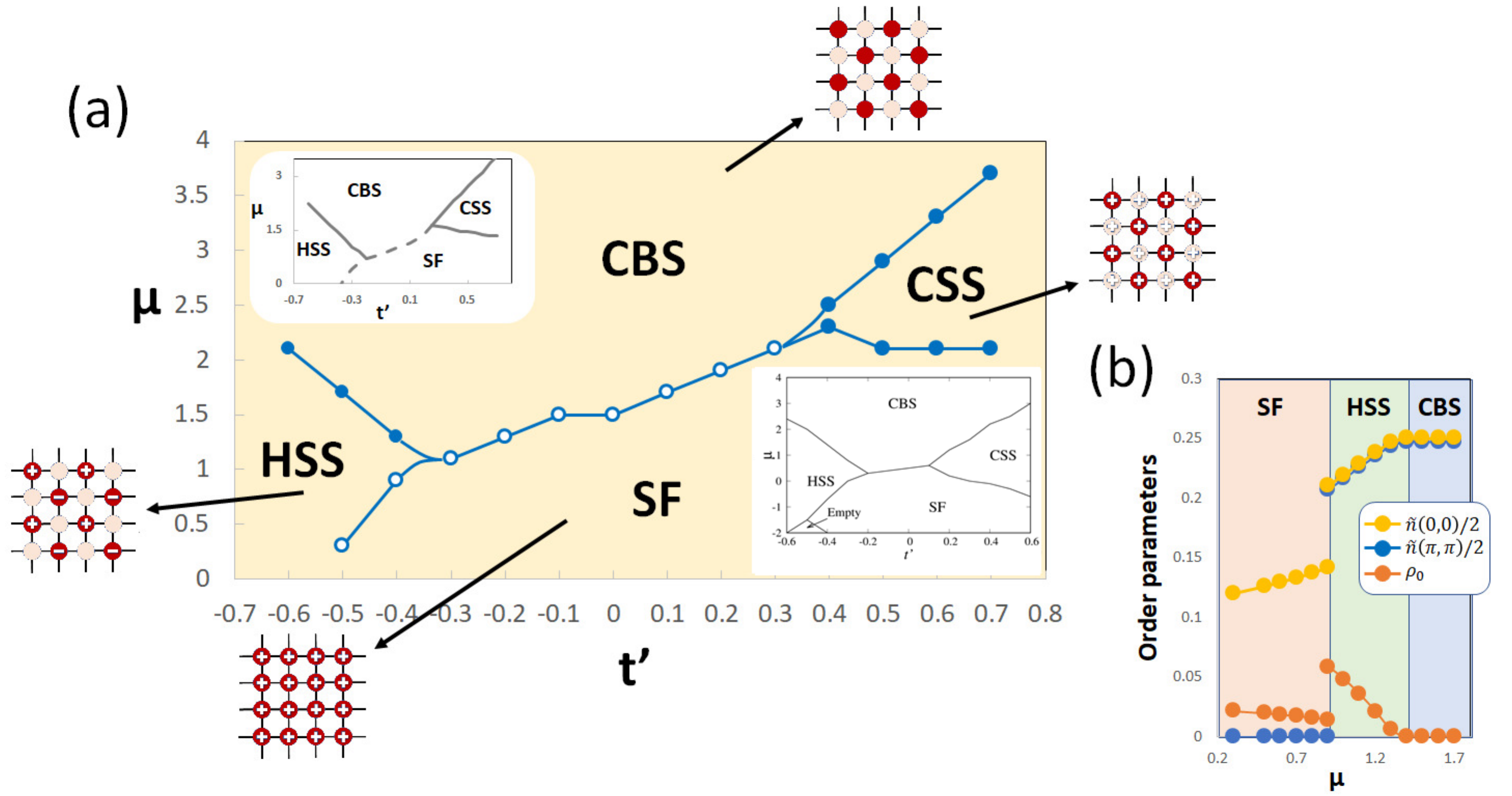}
\caption{\raggedright (a) Phase diagram by iPEPS for $(t,V_1,V_2)=(1.0,8.0,0)$. Besides SF, we also have checkerboard solid (CBS), checkerboard supersolid (CSS), and half supersolid (HSS). Filled (empty) circles stand for boundary of second-(first-)order phase transition. 
Curves connecting circles are only guides to eyes and error bars are smaller than the size of symbols. We adopt $D=8$ results near the boundaries for better estimating the transition points. 
Rules above apply for every phase diagram in the following content. 
Upper and lower inset are diagrams from ref. \onlinecite{ChenS} and our mean-field theory under the same conditions. 
In the upper inset, solid (dotted) curves are boundaries for second-(first-)order transition. 
We demonstrate the schematic real-space structures of our states next to each phase and their definitions for each symbol are the same as those in Fig. \ref{Fig.1}. 
Note that the sign in the circles corresponds to $\arg[\langle b_i \rangle]$ and a site with no sign indication means its $\langle b_{i} \rangle$=0. 
(b) Variations of order parameters for $t'=-0.4$ cut. In (a) and (b) $t=1$ is taken to be the energy unit.}
\label{Fig.2}
\end{figure*}

Since we focus on phases for the filling in $0\le \langle n_i \rangle \le 1$ with the averaged particle density $n_0 \le 0.5$, we can only have three different crystal configurations: checkerboard (CB), stripe, and quarter-filled (QF) modulations. 
Here, $\langle n_i \rangle$ means the particle density on site $i$.
The configuration of the particles are drawn in Figs. \ref{Fig.1}(a)--\ref{Fig.1}(c). 
For the solid phases, CB and stripe contain two filled sites while QF only has one. These solids emerge at commensurate fillings for the $2 \times 2$ unit cell of iPEPS. 
Any underlying SS states are claimed to exist by hole doping or particle doping in these three perfect crystals \cite{ChenSS}. 
To systematically distinguish these three solid phases, we calculate the static structure factor,
\begin{equation}
\begin{aligned}
\tilde{n}(\textbf{k})=\frac{1}{N_C}\sum_{i\in C}\langle n_{i} \rangle e^{i\textbf{k}\cdot \textbf{r}_i},
\end{aligned}
\label{DW}
\end{equation}
where $C$ means unit cell and therefore $N_C$ is equal to 4 for iPEPS. 
$\textbf{r}_i$ is the coordinate of location for each site. 
Depending on the patterns of these three solid phases, $\tilde{n}(\textbf{k})$ can be non-zero at $\textbf{k}=(0,0)$, $(0,\pi)$, $(\pi,0)$, or $(\pi,\pi)$. 
The finite value of $\tilde{n}(\pi,\pi)$ or $\tilde{n}(0,\pi)$ ($\tilde{n}(\pi,0)$) relates to the presence of CB or stripe orders, respectively. 
$\tilde{n}(0,0)$ represents the averaged particle density. Thus, $\tilde{n}(0,0)$ is always non-zero in the following calculations.
In order to distinguish those three configurations in Figs. \ref{Fig.1}(a)--\ref{Fig.1}(c), we apply the same definition encoded in ref. \onlinecite{ChenSS}. 
That is, for a state to be of CB pattern, we demand that its $\tilde{n}(\pi,\pi)\neq 0$ and $\tilde{n}(\pi,0)$, $\tilde{n}(0,\pi)=0$. 
For the rest two orders, the stripe phase fulfills the condition $|\tilde{n}(\pi,0)-\tilde{n}(0,\pi)|\neq 0$ and $\tilde{n}(\pi,\pi)=0$, while the QF phase satisfies $|\tilde{n}(\pi,0)-\tilde{n}(0,\pi)| = 0$ under that $\tilde{n}(\pi,0)$ and $\tilde{n}(0,\pi)\neq 0$.

Besides the structural order of particle density, to see if a phase is SF/SS, one needs to examine its condensate density $\rho_0$, which is characterized by the non-zero value of the off-diagonal components. In this paper, we simply define this order parameter as the following:
\begin{equation}
\begin{aligned}
\rho_0= \frac{1}{N_C}\sum_{i\in C} | \langle b_{i} \rangle |^2.
\end{aligned}
\label{SF}
\end{equation}
This reflects only the net SF density. 
In addition, we evaluate the sign of $\langle b_i \rangle$, namely $\arg[\langle b_i \rangle]$, for each site, because we expect the frustration between $t$ and $t'$ to generate various patterns of $\arg[\langle b_i \rangle].$
Notice that the Fourier transform for SF density is not performed because usually the modulation of $|\langle b_{i} \rangle|$ is related to the order of particle density. 
Moreover, because of the frustration, $\langle b_{i} \rangle$ may suffer from a sign difference in contrast to nearby sites and therefore, the Fourier transform for a specific $\textbf{k}$ in Eq. (\ref{DW}) might fail to reflect the true situation when a detailed analysis is insufficient. 
As a result, our strategy is to see whether or not the condensate density (Eq. (\ref{SF})) can coexist with the solid order defined in Eq. (\ref{DW}) and calculate $\langle b_{i} \rangle$ for each site in the unit cell for further categorizing each SS phase. 
Modulations of the SF density with different momenta are revealed by ED in Sec. II C. Configurations of possible patterns for arg[$b_i$] are shown in Figs. \ref{Fig.1}(d)--\ref{Fig.1}(g), named after homogeneous, diagonal, collinear, and star sign patterns, respectively. 
They will be further discussed in Sec. II D.

\begin{figure*}[t]
\centering
\includegraphics[width=1.0 \textwidth]{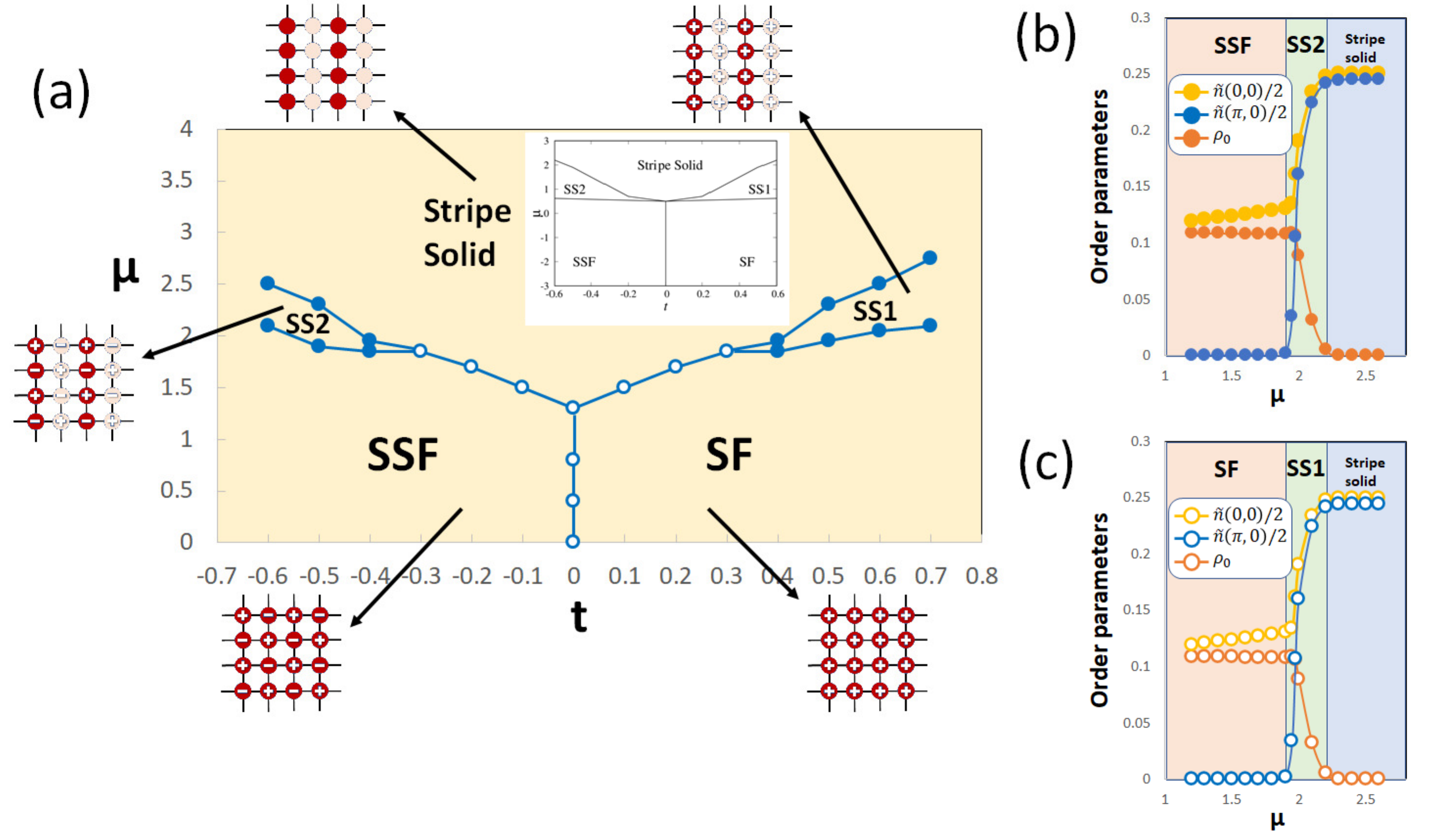}
\caption{\raggedright (a) Phase diagram by iPEPS for $(t',V_1,V_2)=(1.0,0,8.0)$. Now we have two superfluid state: SF and staggered superfluid (SSF), along with two supersolids (SS1, SS2), and stripe solid phases. Upper inset shows the phase diagram from our mean-field theory under the same conditions. Structures of states in real space are shown next to each phase under the same criteria as Fig. \ref{Fig.2}(a). Variations of order parameters for (b) $t=-0.5$ and (c) $t=0.5$ cuts are shown next to the phase diagram. Here, $t'=1$ is taken to be the energy unit.}
\label{Fig.3}
\end{figure*} 

\subsection{\label{sec:level2}Phase diagram construction}
In this section, we strive to construct a couple of phase diagrams in two extremely different component sets of the Hamiltonian (\ref{Hamiltonian}), using iPEPS and mean-field theory. 
In the following phase diagrams obtained by iPEPS, we adopt a fixed bond dimension to be equal to 8. 
This choice already provides a good estimate for the phase diagrams. 
We will discuss the influence of bond dimension $D$ in the Appendix \cite{Wessel}.
Also, we employ the mean-field theory \cite{Matsubara} for the Hamiltonian (\ref{Hamiltonian}) and construct the phase diagrams for the $4\times4$ unit cell at a very low temperature. 
The obtained results are extrapolated to $T=0$ for comparing with the iPEPS results.

\subsubsection{\label{sec:level3}$t=1$, $V_1=8$, $V_2=0$}
First we consider the case for $(t,V_1,V_2)=(1.0,8.0,0)$ while varying $t'$ and $\mu$. 
This is the same scenario discussed in ref. \onlinecite{ChenS}, where the cluster mean-field approach is applied. 
We show our results in Fig. \ref{Fig.2}(a) together with theirs.

In Fig. \ref{Fig.2}(a), the phase boundaries denoted by filled (empty) blue circles indicate the second-(first-)order phase transition. 
iPEPS is able to capture the first-order phase transitions quantitatively.
When the computations are started near the first-order transition with different initial states, each energy converges at a different value. 
By detecting the energy cross, we get to decide the transition points, as discussed in ref. \onlinecite{Corboz3}. 
On the other hand, for the second-order phase transitions, orders can be formed continuously near the boundaries. 
To determine the first-order transition points and evaluate relative errors, we perform the calculation along a vertical cut with fixed $t'$ and slice the value of $\mu$ in the unit of 0.1. 
If we find a energy level crossing between consecutive $\mu=x$ and $\mu=x+0.1$, we then adopt $\mu=x+0.05$ as the transition point, with error equal to 0.1. 
For the second-order phase transition, we set $\langle O \rangle \neq 0$ if its value is larger than $10^{-2}$. 
We define the second-order transition point as the midpoint of two consecutive $\mu$, where the order parameter becomes nonzero.
Thus, the error of the second-order transition point is also equal to 0.1.

The upper inset in Fig. \ref{Fig.2}(a) is the result presented in ref. \onlinecite{ChenS}, where the solid (dotted) curves represent the second-(first-)order transition. 
Our mean-field phase diagram is also shown in the lower inset. 
One can clearly see that these phase diagrams coming from different methods qualitatively agree with each other.
We notice that the mean-field phase boundaries deviate from those of the iPEPS calculation. However, in Ref. \onlinecite{ChenS}, they have shown that after scaling, phase boundaries move upward and become closer to the iPEPS results.
Thus, the mean-field calculation also captures the right phases. In fact, the mean-field analysis can be helpful in understanding the background causes of each phase, which will be discussed in Sec. II D.

Two largest phases in Fig. \ref{Fig.2}(a) are the SF and CBS phases, separated by a first-order boundary and two SS phases. 
If a direct transition between SF and CBS takes place, the transition is of first order due to the breaking of different symmetries. 
In contrast, if CSS stays in between these two phases, the second-order transition is allowed at the phase boundary between SF and CSS, because it is characterized by the emergence of CBS that breaks the $Z_2$ symmetry. 
Thus, we expect that the criticality of this phase transition is explained by the Ising universality. 
The CSS-CBS transition belongs to the SF-insulator transition class \cite{FisherS}. As a result, it is also continuous.

\begin{figure}[h!]
\includegraphics[width=0.5 \textwidth]{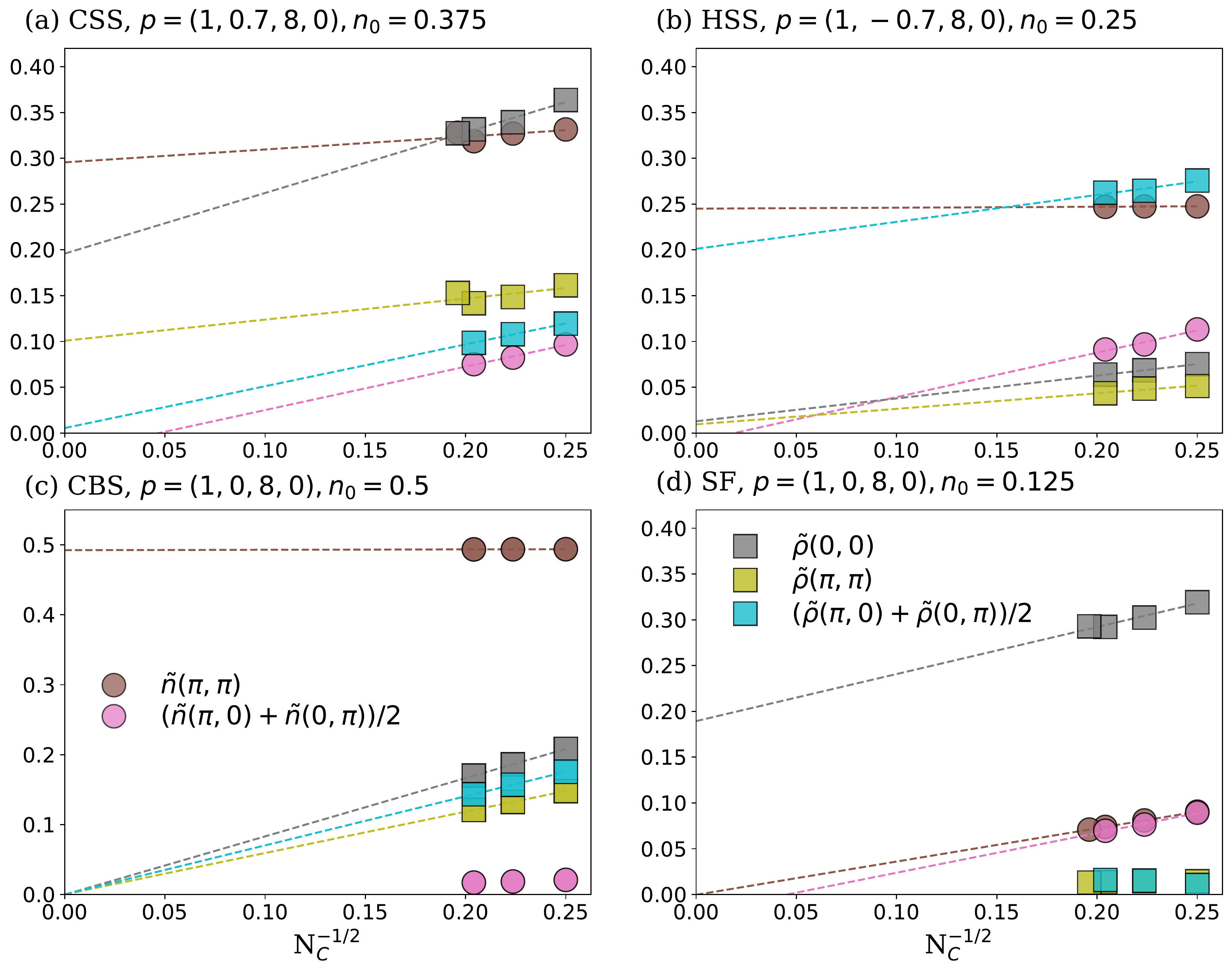}
\caption{\raggedright Finite-size scaling from ED calculation on four underlying states: (a)CSS, (b)HSS, (c)CBS, and (d)SF, within the phase diagram in Fig. ~\ref{Fig.2}(a). Parameter set-up $p=(t,t^\prime,V_1,V_2)$ and averaged particle filling $n_0$ are indicated as the figure titles. Interpolation is applied between two particle numbers closest to our desired filling when such particle number is incommensurate to the lattice size. Dashed lines are fittings to the scaling formula $O(N_C) = O_\infty+\alpha/\sqrt{N_C}$ for the non-vanishing orders.}
\label{Fig.ED_1}

\end{figure} 
\begin{figure}[h!]
\includegraphics[width=0.5 \textwidth]{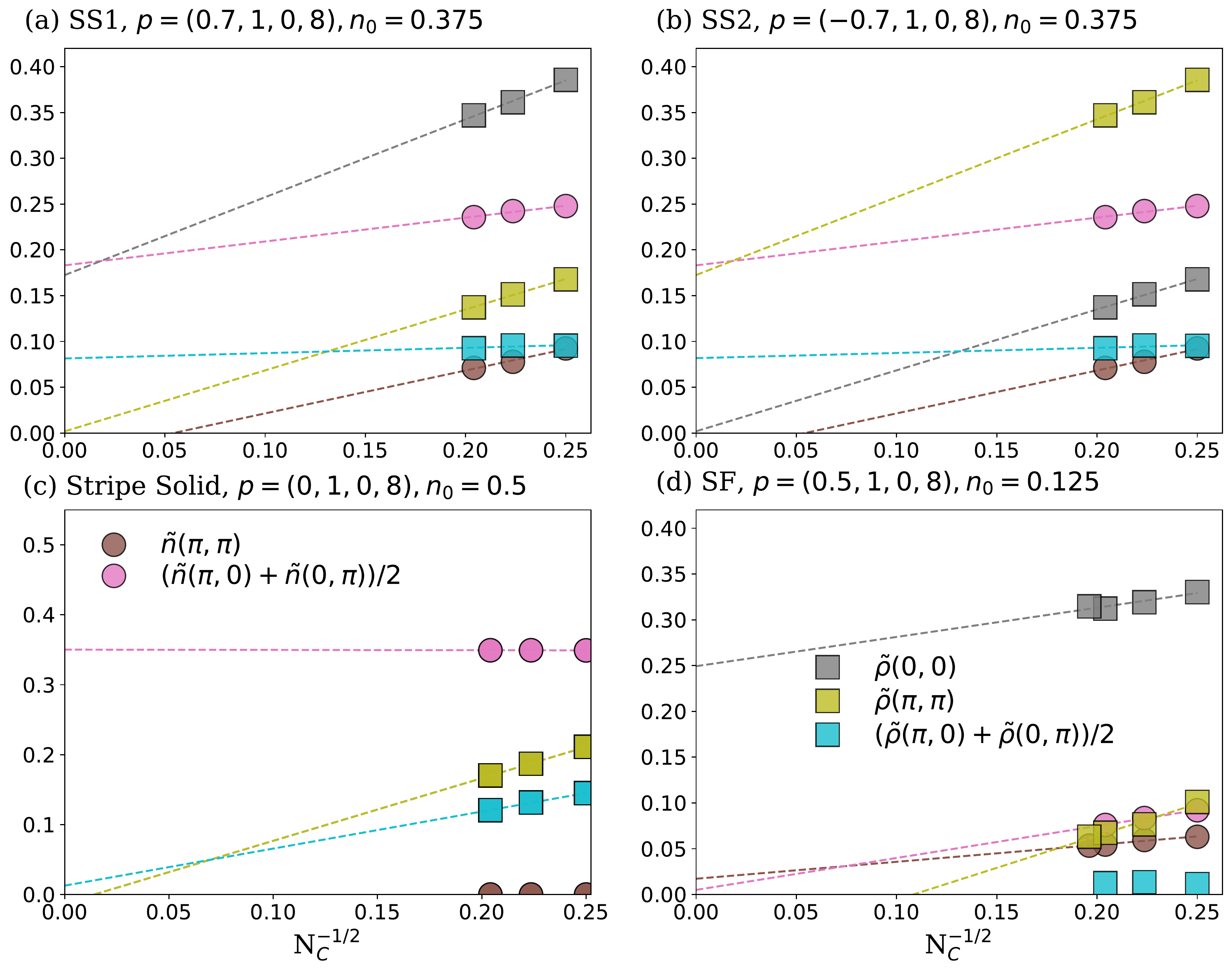}
\caption{\raggedright Same figures as those in Fig. \ref{Fig.ED_1} for (a)SS1, (b)SS2, (c)Stripe solid, and (d)SF within the phase diagram in Fig. ~\ref{Fig.3}(a). Note that in (c), yellow and grey squares overlap completely. This is due to the relation of orders at $t=0$ discussed in the main text.}
\label{Fig.ED_2}
\end{figure}

Among all the phases in Fig. \ref{Fig.2}(a), HSS is considered to be one of the exotic phases induced by frustrated hopping term \cite{Dong, ChenS}. 
This frustration is unavoidable because of the negative $t'$. One of the characteristic feature of HSS is that the staggered pattern of $\arg[\langle b_i \rangle]$ appears along the nnn bonds.
We demonstrate the phase transitions from SF to HSS and from HSS to CBS for $t'=-0.4$ in Fig. \ref{Fig.2}(b). 
A clear first-order transition can be seen at the phase boundary between SF and HSS, while the continuous transition occurs at the HSS-CBS transition. 
This is because at the SF-HSS transition, there is a discontinuous sign change of $\langle b_{i} \rangle$ from $\textbf{++}$ to $\textbf{+}$ $\textbf{--}$ in nearby sites. 
Therefore, the phase transition must be of first order. 
Since the superfluidity disappears at the HSS-CBS transition, its universality belongs to the SF-insulator class and therefore the second-order transition occurs.

\begin{figure*}[t]
\centering
\includegraphics[width=1.0 \textwidth]{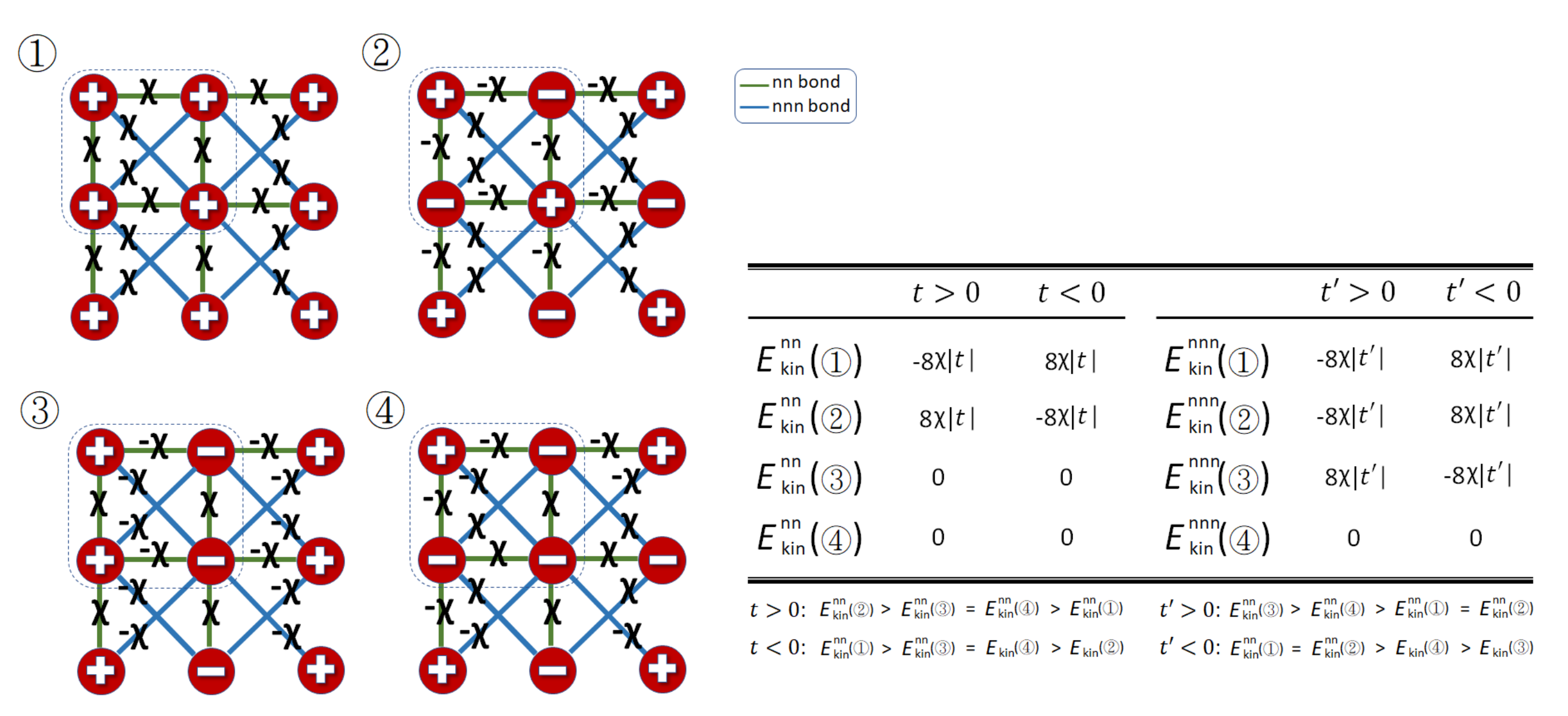}
\caption{\raggedright \textcircled{1}-\textcircled{4} Hopping configurations within 2 by 2 unit cell, corresponding to homogeneous, diagonal, collinear, and star sign patterns, respectively. The sign in the red circle represents $\arg[\langle b_i \rangle]$. Dashed squares enclose the area of unit cells. The total nn (nnn) hopping energies for each configuration are shown in the table. For $t>0$, the most stable configuration is \textcircled{1}, while it becomes configuration \textcircled{2} for $t<0$. As for $t'$, if it stays to be positive then configurations \textcircled{1} and \textcircled{2} are equally favored but as $t'<0$, configuration \textcircled{3} becomes the most stable.}
\label{Fig.4}
\end{figure*} 

\subsubsection{\label{sec:level3}$t'=1$, $V_1=0$, $V_2=8$}
Now let us consider another completely different scenario. We set $(t',V_1,V_2)=(1.0,0,8.0)$ while varying $t$ and $\mu$. 
The phase diagram is shown in Fig. \ref{Fig.3}(a). 
As shown in the inset of Fig. \ref{Fig.3}(a), the mean-field theory qualitatively explains the phase diagram obtained by iPEPS. 
The solid state now has a stripe-like modulation, which is due to the nnn repulsive interaction $(V_2)$. 
For superfluid states, we now have two different states, the normal superfluid (SF) and staggered superfluid (SSF) states. 
Although we can remove the frustration among the hopping terms thanks to the positive $t'$, the staggered pattern of $\arg[\langle b_i \rangle]$ appears.
In $|t|\gtrsim 0.4$, they will evolve into two separate SS states continuously by increasing $\mu$, and end up into the stripe solid phase.

In Figs. \ref{Fig.3}(b) and \ref{Fig.3}(c), we demonstrate the variations of order parameters for superfluid $\to$ supersolid $\to$ solid phases. 
The phase transition from SF(SSF) to SS1(SS2) is explained by the emergence of the stripe solid order. 
This criticality is expected to associate with the three-dimensional Ashkin-Teller (AT) model, where the weak first-order transition or second-order transition explained by the Ising universality class takes place \cite{Musial}. 
In the present iPEPS calculation, the transition seems to be continuous. 
However, we need to investigate more computations near the phase boundaries for making a decisive claim, probably with full-update technique or iPEPS with variational optimization. Further researches will be considered in future works. 
The phase boundary between SS1(SS2) and the stripe solid is of continuous transition because the SF-insulator \cite{FisherS} type transition is expected. 
Combining with the observation upon Fig. \ref{Fig.2}(a), we conclude the following two rules for phase transition: (I) Direct SF $\to$ solid transition is of first order, because of the breaking of different symmetries, according to the Ginzburg-Landau-Wilson paradigm. 
However, if we have SS sandwiched in between SF and solid, then continuous symmetry breaking can happen. 
And (II) if there is a local phase change of $\langle b_{i} \rangle$, then the phase transition falls into the first order owing to the absence of a local $Z_2$ gauge in our model. 
One last point to be noted is that Figs. \ref{Fig.3}(b) and \ref{Fig.3}(c) look nearly identical.
This reflects the symmetry of phase diagram under positive and negative $t$. 
The reason will be discussed in the next section.

\begin{figure*}[t]
\centering
\includegraphics[width=1.0 \textwidth]{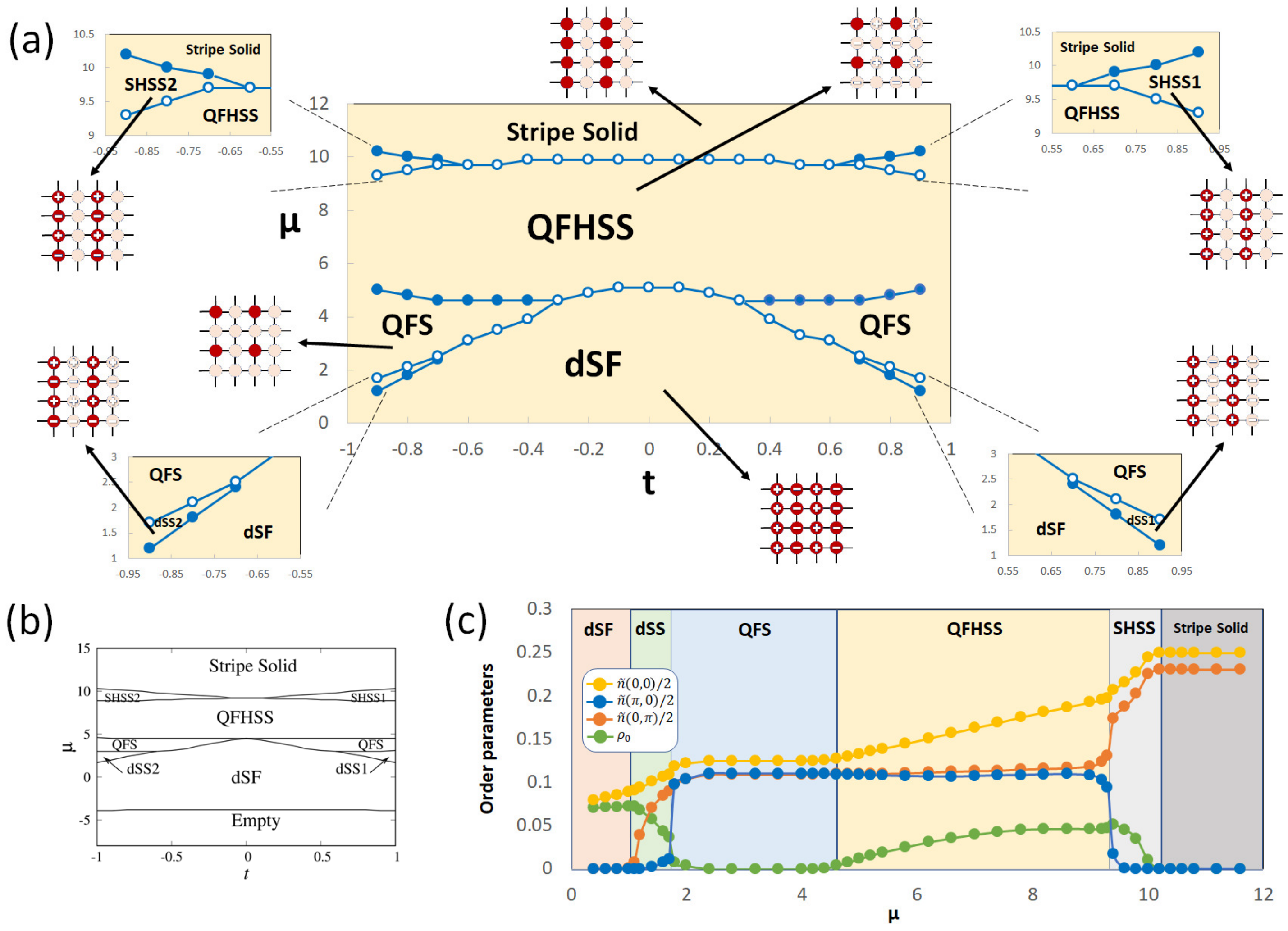}
\caption{\raggedright (a) Phase diagram by iPEPS for $(t',V_1,V_2)=(-1.0,4.0,4.0)$. Our superfluid phase now has a $d$-wave symmetry (dSF). When $|t|$ is small, dSF enters the QF half supersolid (QFHSS) phase through a first-order phase transition; while when $|t|$ is large enough, QF solid (QFS) is sandwiched between QFHSS and dSF. All phases end up in the stripe solid phase when $\mu$ is large enough. We also have four narrow phases in between stripe solid/QFHSS and QFS/dSF when $|t|$ is large. These four phases contain two stripe half supersolid (SHSS1/SHSS2) and two $d$-wave supersolid (dSS1/dSS2), indicated in the enlarged phase diagram sectors. (b) Phase diagram from the mean-field theory under the same conditions. (c) Variations of order parameters for $t=0.9$ cut. Here, $|t'|=1$ is taken to be the energy unit.}
\label{Fig.5}
\end{figure*} 

\subsection{\label{sec:level2}Exact diagonalization}

In this section, we present results obtained from the study of our complementary ED calculation. 
Due to the fact that the finite-size effect becomes influential near the phase boundaries, instead of demonstrating the phase transitions, we focus on providing further evidences of the existence for each of the nine phases in Figs. \ref{Fig.2}(a) and \ref{Fig.3}(a). 
We pick one set of system parameters, $p\equiv (t,t^\prime,V_1,V_2)$, and particle density $n_0$ for each phase. 
To evaluate the values in the thermodynamic limit, we perform the finite-size analysis by using the scaling formula $O(N_C) = O_\infty + \alpha/\sqrt{N_C}$. 
Cluster sizes used here are mainly $N_C = 16$, $20$, and $24$. 
When no stripe-like order is expected, we include the results up to the 26-site cluster in the finite-size analysis. Note that the 26-site cluster used here \cite{jaklivc2000finite} is not compatible with the stripe order under the periodic boundary condition.

Since translational symmetry breaking can not be seen directly in the ED ground states, we examine the presence of the long-range order from the correlation functions. 
For the particle and SF density, we calculate their structural orders defined by
\begin{align}
S_n(\mathbf{k}) &= \frac{1}{(N_C)^2} \sum_{i,j\in C}\langle n_in_j\rangle e^{i\mathbf{k}\cdot(\mathbf{r}_j-\mathbf{r}_i)}
\end{align}
and
\begin{align}
S_\rho(\mathbf{k}) &= \frac{1}{(N_C)^2} \sum_{i,j\in C}\langle b_i^\dagger b_j\rangle e^{i\mathbf{k}\cdot(\mathbf{r}_j-\mathbf{r}_i)},
\end{align}
respectively. 
We then define the order parameters by taking the square roots: $\tilde{n}(\mathbf{k}) = \sqrt{|S_n(\mathbf{k})|}$ and $\tilde{\rho}(\mathbf{k}) = \sqrt{|S_\rho(\mathbf{k})|}$.
Therefore, CB and stripe orders are represented by $\tilde{n}(\pi,\pi)$ and $(\tilde{n}(\pi,0)+\tilde{n}(0,\pi))/2$, respectively.

The results for the finite-size scaling plot for $(t, V_1,V_2) = (1,8,0)$ and $(t', V_1,V_2) = (1,0,8)$ are shown in Figs. \ref{Fig.ED_1} and \ref{Fig.ED_2}, respectively. 
For $(t', V_1,V_2) = (1,0,8)$, we notice that $H(\pm t,t',V_1,V_2)$ are related by the unitary transformation of $Tb_iT^{-1}= (-1)^{r_{ix}+r_{iy}}b_i$, which leaves every parameter unchanged but inverts the sign of $t$ in the Hamiltonian. Notice that this transformation cannot remove the frustration induced by the kinetic terms and we will discuss this issue in the next section. Using the above unitary transformation, we can infer the results of desired orders as $t\rightarrow -t$, by mapping $\tilde{\rho}(0,0)\rightarrow \tilde{\rho}(\pi,\pi)$ and $\tilde{\rho}(\pi,\pi)\rightarrow \tilde{\rho}(0,0)$, while every other order stays the same. 
Such relation can be seen between SS1 in Fig.~\ref{Fig.ED_2}(a) and SS2 in Fig.~\ref{Fig.ED_2}(b). 
This also accounts for the equivalence of Figs. \ref{Fig.3}(b) and \ref{Fig.3}(c), and that of $\tilde{\rho}(0,0)$ and $\tilde{\rho}(\pi,\pi)$ in Fig.~\ref{Fig.ED_2}(c), where $t=0$ is chosen. 
Thus, we only show the results for the four states except SSF in Fig. \ref{Fig.ED_2}, because the outcomes for SSF are equivalent to those of SF by the above-mentioned transformation.

Figure \ref{Fig.ED_1} shows the cluster-size-dependence of the structural order parameters for four sets of $p$ and $n_0$. 
In Figs. \ref{Fig.ED_1}(a) and \ref{Fig.ED_1}(b), we demonstrate two SS states whose order parameters for the particle density show non-zero CB order ($\tilde{n}(\pi,\pi)$). 
In Fig. \ref{Fig.ED_1}(a), the SF density possesses the same modulation as that of the particle density in CSS. 
In contrast, the SF density for $(\tilde{\rho}(\pi,0)+\tilde{\rho}(0,\pi))/2$ becomes non-zero in HSS, as shown in Fig. \ref{Fig.ED_1}(b).
This is due to the reason that the pattern of arg[$\langle b_i \rangle$] in HSS is classified into the collinear type in Fig. \ref{Fig.1}(f). 
In the CB solid phase, all SF density orders are extrapolated to zero and only $\tilde{n}(\pi,\pi)$ remains, as shown in Fig. \ref{Fig.ED_1}(c).
Another state with only one existing order is the SF state in Fig. \ref{Fig.ED_1}(d), where all but $\tilde{\rho}(0,0)$ are extrapolated to values either close to or below zero.

As mentioned above, Figs. \ref{Fig.ED_2}(a) and \ref{Fig.ED_2}(b) are symmetrical after interchanging $\tilde{\rho}(0,0)$ and $\tilde{\rho}(\pi,\pi)$ orders. 
Non-zero $(\tilde{n}(\pi,0)+\tilde{n}(0,\pi))/2$ order also indicates that they are both SS states accompanied by the stripe solid order.
The result for the stripe solid is shown in Figs. \ref{Fig.ED_2}(c), where the only remaining order is $(\tilde{n}(\pi,0)+\tilde{n}(0,\pi))/2$.
In Fig. \ref{Fig.ED_2}(d), only $\tilde{\rho}(0,0)$ survives in the thermodynamic limit, which means that the SF phase appears. 
We conclude that the results by ED are consistent with those by iPEPS and mean-field theory.

\subsection{\label{sec:level2}Discussion}

\subsubsection{\label{sec:level3}Formation of SF/SS states}
So far, we have demonstrated several SF/SS phases, besides the CSS and HSS which were found earlier \cite{Dong, ChenS}. 
In this section, we try to clarify the mechanism of the emergence of CSS and HSS, and then make a simple categorization. 
To begin with, we have learned that introducing the diagonal nnn terms into the Hamiltonian (\ref{Hamiltonian}) is necessary for the formation of SS \cite{Batrouni, Hebert}. 
The CB (stripe) pattern is caused by the dominant $V_1$ ($V_2$) interaction, because bosons tend to occupy the nnn (nn) sites in avoidance of repulsive interaction. 
When $V_1$ competes with $V_2$, the third structural order, namely QF order, can be formed \cite{ChenSS}. 
We will demonstrate the QF order in the latter section.

What interests us the most is the phase modulation of SF density on each site, leading to the exotic SF/SS phases. 
In order to understand the underlying causes, we first decouple our hopping observable through a mean-field treatment:
\begin{equation}
\begin{aligned}
&b_{i}=\langle b_{i} \rangle+(b_i-\langle b_{i} \rangle)=\langle b_{i} \rangle+\delta b_i ,
\end{aligned}
\label{MF1}
\end{equation} 
and then our hopping value can be approximated as:
\begin{equation}
\begin{aligned}
\chi_{ij} &\equiv \langle b^{\dagger}_{i} b_{j} \rangle \\
&= \langle b^{\dagger}_{i} \rangle \langle b_{j} \rangle +\langle \delta b^{\dagger}_{i} \delta b_{j} \rangle \thickapprox \langle b^{\dagger}_{i} \rangle \langle b_{j} \rangle.
\end{aligned}
\label{MF2}
\end{equation} 
The approximation in Eq. (\ref{MF2}) holds as away from the phase boundaries. By decoupling the hopping term in the mean-field way, we are now more able to look inside what happens for different $t$ and $t'$. 

Because the SS state always starts from SF by enhancing the chemical potential (filling), we shall first analyze various SF states. 
In Figs. \ref{Fig.1}(d)--\ref{Fig.1}(g), we present four possible SF states with different patterns of on-site arg[$\langle b_{i} \rangle$]. 
Now we start our iPEPS calculation with an initial state of certain filling and a hopping strength.
For the simplicity, we focus on the sign of $|\langle b^{\dagger}_{i} b_{j}+H.C. \rangle| \equiv \chi_{i,j}$ for $(i,j) \in \langle ij \rangle$ and $\langle \langle ij \rangle \rangle$ and suppose that all $\chi_{i,j}$ are homogeneous.
From Eq. (\ref{MF2}), we can compose the pattern of $\chi_{i,j}$ on each bond in the unit cell, as shown in Figs. \ref{Fig.4}\textcircled{1}--\ref{Fig.4}\textcircled{4}, corresponding to homogeneous, diagonal, collinear, and star sign patterns, as indicated in Figs. \ref{Fig.1}(d)--\ref{Fig.1}(g).
The overall nn (nnn) kinetic energies after multiplied by $t$ ($t'$) are shown in the table of Fig. \ref{Fig.4}.

One can easily find out that in $t>0$ and $t'>0$, the configuration with the lowest kinetic energy is homogeneous pattern, meaning that no frustration happens. 
In $t>0$ and $t'<0$, the total nn hopping energy $E^{\text{nn}}_{\text{kin}}$ still prefers the homogeneous pattern; $E^{\text{nnn}}_{\text{kin}}$, however, is in favor of the collinear pattern.
In this way, the frustration between $E^{\text{nn}}_{\text{kin}}$ and $E^{\text{nnn}}_{\text{kin}}$ realizes the HSS phase in Fig. \ref{Fig.2}.

As shown in Fig. \ref{Fig.3}, when $t'>0$, homogeneous and diagonal patterns are equally favored for $t=0$, where the phase boundary between SF and SSF exists.
In fact, the configuration with the lowest $E^{\text{nn}}_{\text{kin}}$ is selected by the sign of $t$. 
The condition of $t>0$ prefers the homogeneous pattern of Fig. \ref{Fig.4}\textcircled{1}, while $t<0$ prefers the diagonal pattern of Fig. \ref{Fig.4}\textcircled{2}.
This results in the reduction of frustration between the kinetic terms. 
In the same manner, two SS states stabilized by $V_2$ can be formed with the different patterns of on-site arg[$\langle b_{i} \rangle$].

The rules for constructing desired SF/SS phases are summarized as follows:

\vspace{0.1cm}
{\raggedright(I) Next-nearest-neighbor terms are needed for SS.}

\vspace{0.1cm}
{\raggedright(II) Dominant $V_1$ ($V_2$) favors CB (stripe) pattern. QF order can be formed under competing $V_1$ and $V_2$.}

\vspace{0.1cm}
{\raggedright(III) In $t>0$, homogeneous pattern of $\arg[\langle b_i \rangle]$ is favored as $t'>0$, while it competes with collinear pattern of $\arg[\langle b_i \rangle]$ as $t'<0$.}

\vspace{0.1cm}
{\raggedright(IV) In $t<0$, diagonal pattern of $\arg[\langle b_i \rangle]$ is favored as $t'>0$, while it competes with collinear pattern of $\arg[\langle b_i \rangle]$ as $t'<0$.}

\vspace{0.1cm}

One fact revealed by the above rules is that the configuration of star sign pattern has no chance to be seen for the isotropic EBH model. 
With these four rules, we are now able to access desired phases by manipulating each term of the EBH. 
We first introduce the nnn terms for SS (Rule (I)) and count on $V_1$/$V_2$ for deciding the leading real-space order (Rule(II)). 
As for the pattern of $\arg[\langle b_i \rangle]$, we can have the homogeneous/diagonal sign pattern in $t'>0$ and $t>0$/$t<0$ (Rule(III)/(IV)). 
When the frustration exists in the model, namely $t'<0$, homogeneous sign pattern competes with collinear sign pattern in $t>0$ (Rule(III)), and diagonal sign pattern competes with collinear sign pattern in $t<0$  (Rule(IV)). In the previous section, we have learned that our Hamiltonian (\ref{Hamiltonian}), after a unitary transformation, is unchanged but with a inverted sign of $t$. Here, according to Rule (III) and (IV), we can see that such transformation will leave our model with $t'<0$($t'>0$) to be still (non-)frustrated.

So far, our results fit quite well with the above rules from the mean-field theory. 
But more importantly, we would like to see if these rules can be used to predict the existence of certain SF or SS states. 
Therefore, in the next section, we examine the scenario with the set-up of $(t',V_1,V_2)=(-1.0,4.0,4.0)$.

\subsubsection{\label{sec:level3}$t'=-1$, $V_1=4$, $V_2=4$}

For $t’=-1$, the system includes the frustration between the hopping terms, independent of the sign of $t$. The frustration cannot be removed by the unitary transformation discussed in Sec. II C. Before we look at the results, we conjecture what kinds of states can appear from our rules. 
Because $V_1$ and $V_2$ are competing, we expect that QF order can be generated (Fig. \ref{Fig.1}(c)) \cite{ChenSS, Ng, Dang}. 
Also, we expect that the collinear pattern of $\arg[\langle b_i \rangle]$ is favored in our SF state when $|t'|$ is dominant. 
This provides the so-called $d$-wave superfluid (dSF) \cite{Motrunich}, which shares the same $d$-wave pairing symmetry as the high-$T_c$ superconductivity \cite{Anderson}. 
By increasing $\mu$, we expect the QF solid and QF SS phases before finally entering the stripe solid phase \cite{ChenSS}. 
However, our SS is presumably different from the conventional ones where no frustration is presented.

The resulting phase diagram is shown in Fig. \ref{Fig.5}(a). 
First, iPEPS phase diagram again qualitatively agrees with the one by mean-field theory in Fig. \ref{Fig.5}(b). 
Here we isolate the mean-field phase diagram for better demonstration. 
In our phase diagram, there appear one superfluid state (dSF), five supersolid states (dSS1, dSS2, QFHSS, SHSS1, SHSS2), and two solid states (QFS, stripe solid) in $\mu>0$. 
All our phases start with dSF, which was named after $d$-wave Bose liquid in the previous works \cite{Motrunich, Sheng}. 
But unlike their model, which included a four-site ring interaction, our dSF is realized completely from the two-site terms. 
More importantly, the frustrated hopping terms play a crucial role in both scenarios. 
When $|t|\lesssim 0.3$, dSF enters the QF supersolid phase, which is named after the QF half supersolid (QFHSS), through a first-order transition. 
For $|t|\gtrsim 0.3$, the QF solid (QFS) phase appears in between QFHSS and dSF. This QF solid may help the connection between QFHSS and dSF. 
As $\mu$ increases, QFHSS is replaced by the stripe solid after the first order transition. 
This is because the QF solid order is not simply composed of two perpendicular stripe orders.
The first-order transition from the QF phase to the stripe phase has also been reported in Ref. \onlinecite{NgS}.

Our complex phase diagram also contains the stripe SS phases in several narrow regions, indicated by those four panels in Fig. \ref{Fig.5}(a). 
The difference between dSS1/dSS2 and SHSS1/SHSS2 lies on the following point: the orientation of stripe solid order stays parallel/perpendicular to the orientation of sign modulation of $\arg[\langle b_i \rangle]$. 
This can be understood with again the mean-field analysis. 
Since we have two sublattices, we assign $\langle b_{\rm sub1} \rangle = a \in \mathbb{C}$ and $\langle b_{\rm sub2} \rangle = b \in \mathbb{C}$. 
Then because of the inequality of arithmetic and geometric means:
\begin{equation}
\begin{aligned}
a^2+b^2 \geqslant 2|a||b|,
\end{aligned}
\label{AMGM}
\end{equation} 
in order to lower the total energy, when $t>0$, they tend to possess the same sign among one sublattice sites, leading to dSS1 and SHSS1, and vice versa in $t<0$. 
Phase diagrams in Figs. \ref{Fig.5}(a) and \ref{Fig.5}(b) possess the parity symmetry of $\pm t$ due to the same cause explained in Sec. II C. 
In Fig. \ref{Fig.5}(c), we demonstrate the variation of order parameters for $t=0.9$. 
The sudden changes of the order parameter, indicating first-order transition, are very clear on the boundaries of dSS/QFS and QFHSS/SHSS. 
Due to the emergence of the stripe order, the dSF-dSS1(dSS2) transition is again expected to be in association with the type of 3D AT model \cite{Musial}, followed by the first-order dSS1(dSS2)-QFS transition resulted from the breaking of different symmetries \cite{ChenSS, NgS}. 
The QFS-QFHSS transition is continuous because it belongs to the SF-insulator class \cite{FisherS}. 
The QFHSS-SHSS1(SHSS2) transition is of first-order since it is again the QF-to-stripe transition. 
At last, SHSS1(SHSS2) transits to the stripe solid continuously, once again due to the SF-insulator class.

\section{\label{sec:level1}Conclusion}

In this work, we have presented the exotic SF and SS states out of the EBH model in the hard-core limit and have discussed the background mechanism of generating desired SF/SS states, by manipulating the hopping and interacting strengths of the EBH model. 
From the detailed mean-field interpretation, we have concluded with several rules of the SF/SS formation. 
Thanks to the advance of cold atom experiment, which is now able to tune the effective hopping parameters to be negative or even complex \cite{Eckardt}, our states and unveiled rules can be applied experimentally. 
By a simple mapping, the EBH model can be mapped into XXZ or Heisenberg model, where the frustration of interactive terms becomes important for deciding whether a system should be ferro- or antiferromagnetic. 
Therefore, from the side of hard-core boson, there could be some interesting physics that may help interpret the intriguing quantum magnetic states, including the quantum spin liquid, more deeply. 
This would be one of our future subjects. 

\section{\label{sec:level1}Acknowledgement}

W.-L.T. is supported by Postdoctoral Research Abroad Program, Project No. 108-2917-I-564-007, from Ministry of Science and Technology (MOST) of Taiwan. Special thanks to Japan-Taiwan Exchange Association for its sponsorship of a short-term research activity in 2018.
H.-K.W. is supported by JQI-NSF-PFC (supported by NSF grant PHY-1607611). 
T.S. is supported by the Creation of new functional devices and high-performance materials to support next-generation industries (CDMSI) and Challenge of Basic Science - Exploring Extremes through Multi-scale Simulations (CBSM2) from MEXT, Japan.

\section{\label{sec:level1}Appendix: iPEPS extrapolation}

\begin{figure}[t]
\centering
\includegraphics[width=0.48 \textwidth]{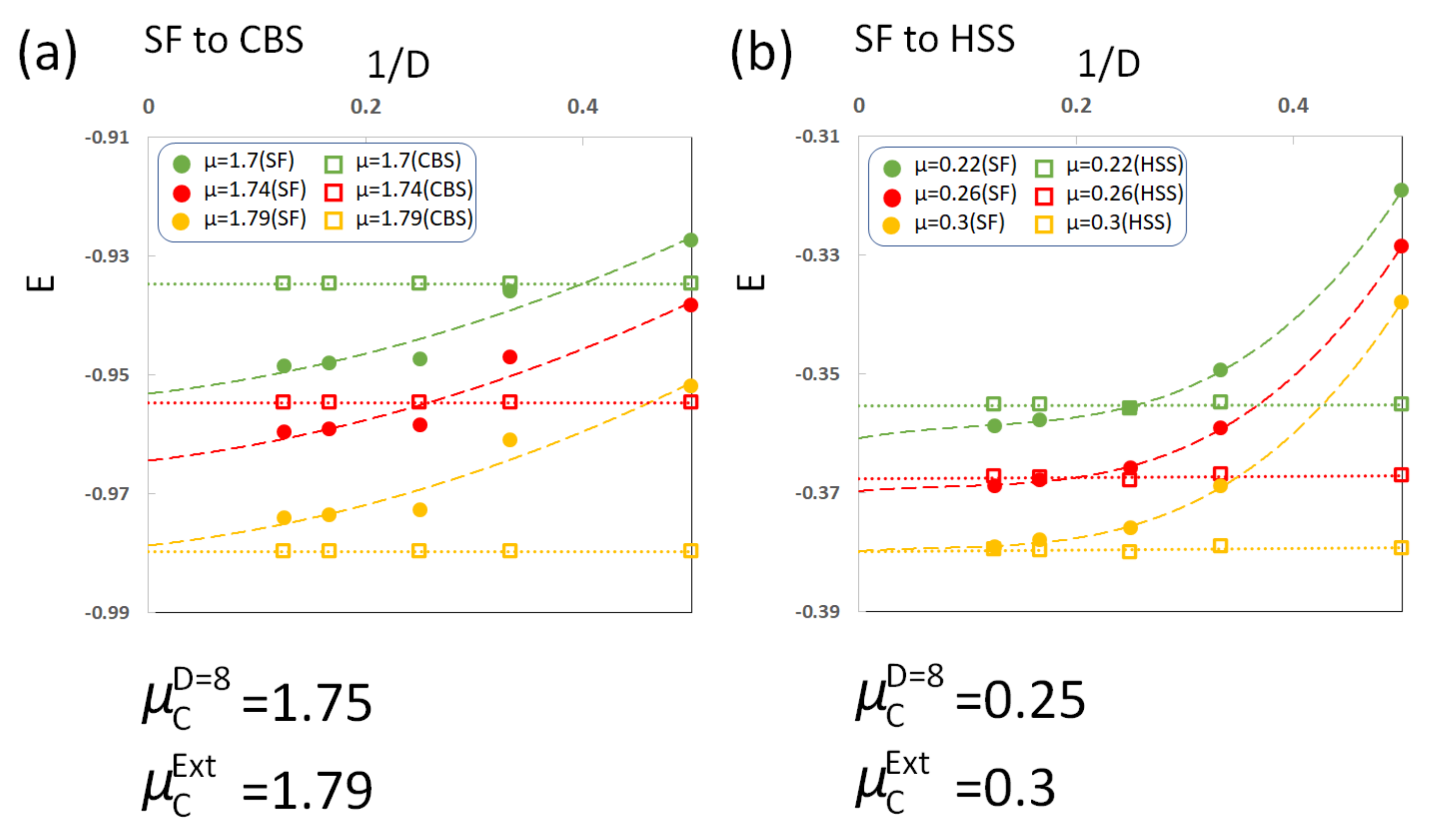}
\caption{\raggedright Extrapolations of energies to infinite $D$ for (a) SF to CBS and (b) SF to HSS phase transitions. We choose $t'=0.1$ for (a) and $t'=-0.5$ for (b). Transition points are determined by comparing extrapolated energies at $\mu^{Ext}_c$. We show the transition points, $\mu^{D=8}_c$ and $\mu^{Ext}_c$, in the lower panel of each figure.}
\label{Fig.a1}
\end{figure} 

Our phase diagrams shown in the main text come from the simple-update iPEPS calculation with fixed bond dimension, $D=8$. 
But since the precision of iPEPS ansatz is highly dependent on the bond dimension, it is required to carefully check its influence to our present model. 
In this Appendix, we extrapolate $D$ to infinity and argue the phase boundaries obtained from the iPEPS calculations. 
Since the phase diagram in Fig. \ref{Fig.2}(a) contains both frustrated and non-frustrated cases, we will focus our studies on those phase boundaries.

\begin{figure}[t]
\centering
\includegraphics[width=0.48 \textwidth]{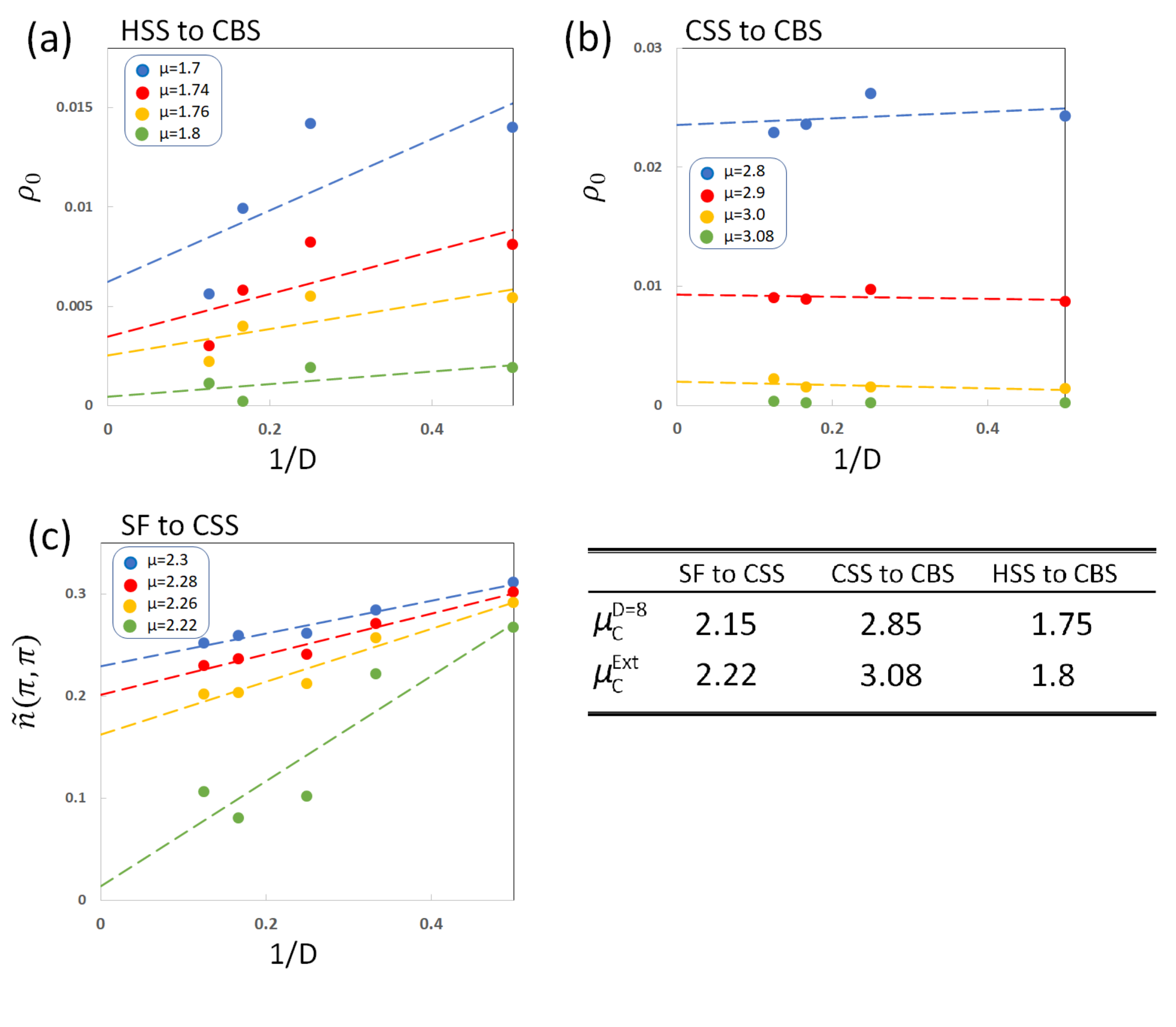}
\caption{\raggedright Extrapolation of order parameters to infinite $D$ for (a) HSS to CBS, (b) CSS to CBS, and (c) SF to CSS phase transitions. We choose $t'=-0.5$ for (a) and $t'=0.5$ for (b) and (c). Transition points are determined as orders are extrapolated to zero or very close to zero. $\mu^{D=8}_c$ and $\mu^{Ext}_c$ are shown in the table for these three phase transitions separately.}
\label{Fig.a2}
\end{figure} 

\subsection{\label{sec:level2}First-order boundaries}

For the phase diagrams with fixed $D$, to determine the first-order phase boundaries, we searched along the vertical cut for consecutive $\mu$ where the energy level crossing takes place for the adjoint phases. 
Therefore, we need to extrapolate the phase energies to infinite $D$ and see whether or not the choices of transition points for our fixed-$D$ phase boundaries are reasonable. 
We then plot the energies for CBS and SF at $t'=0.1$ in Fig. \ref{Fig.a1}(a), and energies for HSS and SF for $t'=-0.5$ in Fig. \ref{Fig.a1}(b). 
We observe that the SF energies seem to vary along with $D$ more largely. 
Therefore, we extrapolate the SF energies with a polynomial fit up to the third order to infinite $D$ and obtain $E_{SF}^{D \to \infty}$. 
On the other hand, energies for HSS and CBS seem to be less affected by the choice of $D$, so we simply choose the values of linear extrapolation as $E_{HSS}^{D \to \infty}$ and $E_{CBS}^{D \to \infty}$. 
We compare the extrapolated values and determine the phase transition point from the condition, $E_{SF}^{D \to \infty}=E_{HSS}^{D \to \infty}$ or $E_{SF}^{D \to \infty}=E_{CBS}^{D \to \infty}$.

As shown in Figs. \ref{Fig.a1}(a) and \ref{Fig.a1}(b), the transition points are located at $\mu^{Ext}_c \approx 1.79$ and $0.3$, respectively, which are around 0.05 larger than the transition points obtained from the $D=8$ calculation. 
This is not surprising, since $E_{SF}^{D=8}$ is only a upper bound of energy and $E_{SF}^{D=8}>E_{SF}^{D \to \infty}$. 
Also, $E_{CBS}$ and $E_{HSS}$ are more stable upon varying $D$. 
Thus, the determined transition points after the extrapolation becomes larger. 
However, since $E_{SF}^{D=8}$ is already a good estimate, the variation of transition points is not obvious.

\subsection{\label{sec:level2}Second-order boundaries}

Similar to the case for the first-order boundaries, we need to extrapolate to infinite $D$ for determining the second-order transition points.
However, the quantity we need here depends on the related order parameters. 
As shown in Fig. \ref{Fig.a2}, we demonstrate the behavior of order parameters as a function of $1/D$ at the phase boundaries of (a) HSS to CBS, (b) CSS to CBS, and (c) SF to CSS. 
We choose $t'$ to be equal to -0.5 for Fig. \ref{Fig.a2}(a) and 0.5 for (b) and (c). 
The values of transition points are put together with previous ones for $D=8$ in the table of Fig. \ref{Fig.a2}. 
One can see that the variation between these two transition points becomes larger for the CSS-CBS transition. 
It is due to the fact that the related order parameter, $\rho_0$, is less sensitive to the different bond dimension. 

\subsection{\label{sec:level2}Extrapolated phase diagram}

\begin{figure}[t]
\centering
\includegraphics[width=0.48 \textwidth]{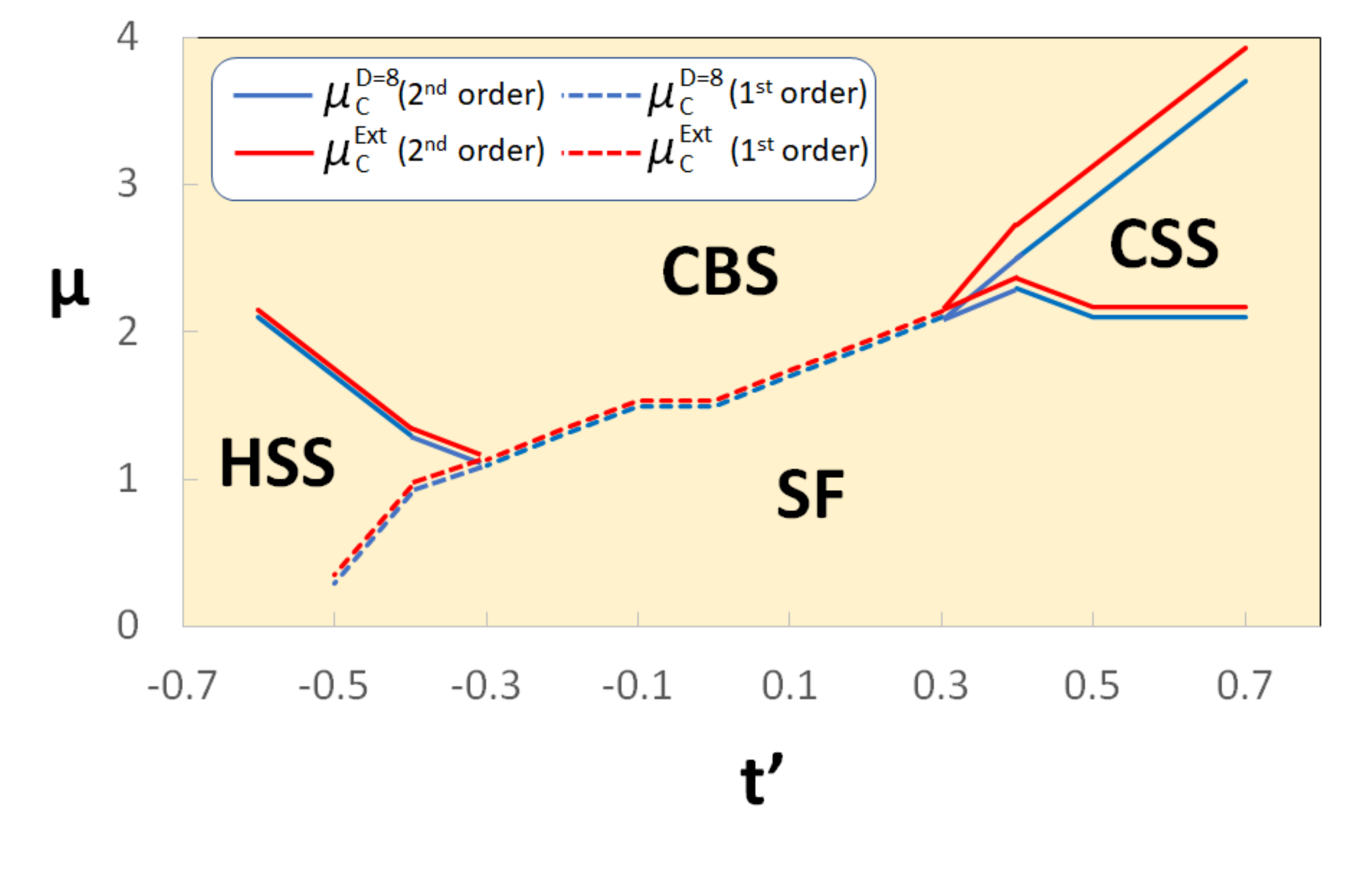}
\caption{\raggedright Phase diagram with transition points determined by fixed-$D$ criteria (blue) and extrapolation (red). Solid (dotted) lines stand for second-(first-)order transition.}
\label{Fig.a3}
\end{figure} 

Since we have already checked the extrapolated transition points along certain cuts for both first- and second-order phase transitions, we would like to reconstruct again the phase diagram and compare it with that of a fixed bond dimension. 
Based on Ref. \cite{Wessel}, we assume that the variation of the transition point is almost independent along the same boundary.
Since we have already sampled one cut for each phase boundary, we shift the boundaries by using the same values obtained in Sec. IV A and Sec. IV B.
The result is shown in Fig. \ref{Fig.a3}. 
One can see that the variation is not very obvious except for the second-order CSS-CBS transition. 
The reason is, as mentioned above, due to the fact that the related order parameter ($\rho_0$) is not very sensitive to the variation of $D$. 
Recall that for fixed-$D$ phase diagram, we determine the second-order transition points from the condition $\langle O \rangle > 10^{-2}$ for consecutive $\mu$. 
If we apply the same standard for the extrapolated phase diagram, then $\mu^{Ext}_c \sim 2.9$, which is much closer to $\mu^{D=8}_c$. 
Therefore, the variation at the CSS-CBS transition point reflects the defect of our previous criteria for determining the transition points. 
Nevertheless, it does not cause any of qualitative change and thus is still acceptable.

%




\bibliography{Hard_Core_Boson_PRR}

\end{document}